\input harvmac
% S-Tables Macro
%
\message{S-Tables Macro v1.0, ACS, TAMU (RANHELP@VENUS.TAMU.EDU)}
%
% Help Text
%
\newhelp\stablestylehelp{You must choose a style between 0 and 3.}%
\newhelp\stablelinehelp{You should not use special hrules when stretching
a table.}%
\newhelp\stablesmultiplehelp{You have tried to place an S-Table inside another
S-Table.  I would recommend not going on.}%
%
% Line Thicknesses (Values)
%
\newdimen\stablesthinline
\stablesthinline=0.4pt
\newdimen\stablesthickline
\stablesthickline=1pt
%
% Border and Internal Line Thicknesses
%
\newif\ifstablesborderthin
\stablesborderthinfalse
\newif\ifstablesinternalthin
\stablesinternalthintrue
\newif\ifstablesomit
\newif\ifstablemode
\newif\ifstablesright
\stablesrightfalse
%
% Save Registers
%
\newdimen\stablesbaselineskip
\newdimen\stableslineskip
\newdimen\stableslineskiplimit
%
% Counters
%
\newcount\stablesmode
\newcount\stableslines
\newcount\stablestemp
\stablestemp=3
\newcount\stablescount
\stablescount=0
\newcount\stableslinet
\stableslinet=0
%
% Table Style Selection
%
% 0 - Centered
% 1 - Left Justified
% 2 - Right Justified
% 3 - Not Justified
%
\newcount\stablestyle
\stablestyle=0
%
% Element Buffering Definitions
%
\def\stablesleft{\quad\hfil}%
\def\stablesright{\hfil\quad}%
%
% Vertical Bar Activation
%
\catcode`\|=\active%
%
% Strut Control
%
\newcount\stablestrutsize
\newbox\stablestrutbox
\setbox\stablestrutbox=\hbox{\vrule height10pt depth5pt width0pt}
\def\stablestrut{\relax\ifmmode%
                         \copy\stablestrutbox%
                       \else%
                         \unhcopy\stablestrutbox%
                       \fi}%
%
% Misc. Internal Stuff
%
\newdimen\stablesborderwidth
\newdimen\stablesinternalwidth
\newdimen\stablesdummy
\newcount\stablesdummyc
\newif\ifstablesin
\stablesinfalse
%
% Table Macros
%
\def\begintable{\stablestart%
  \stablemodetrue%
  \stablesadj%
  \halign%
  \stablesdef}%
\def\stablesadj{%
  \ifcase\stablestyle%
    \hbox to \hsize\bgroup\hss\vbox\bgroup%
  \or%
    \hbox to \hsize\bgroup\vbox\bgroup%
  \or%
    \hbox to \hsize\bgroup\hss\vbox\bgroup%
  \or%
    \hbox\bgroup\vbox\bgroup%
  \else%
    \errhelp=\stablestylehelp%
    \errmessage{Invalid style selected, using default}%
    \hbox to \hsize\bgroup\hss\vbox\bgroup%
  \fi}%
\def\stablesend{\egroup%
  \ifcase\stablestyle%
    \hss\egroup%
  \or%
    \hss\egroup%
  \or%
    \egroup%
  \or%
    \egroup%
  \else%
    \hss\egroup%
  \fi}%
\def\stablestart{%
  \ifstablesin%
    \errhelp=\stablesmultiplehelp%
    \errmessage{An S-Table cannot be placed within an S-Table!}%
  \fi
  \global\stablesintrue%
  \global\advance\stablescount by 1%
  \message{<S-Tables Generating Table \number\stablescount}%
  \begingroup%
  \stablestrutsize=\ht\stablestrutbox%
  \advance\stablestrutsize by \dp\stablestrutbox%
  \ifstablesborderthin%
    \stablesborderwidth=\stablesthinline%
  \else%
    \stablesborderwidth=\stablesthickline%
  \fi%
  \ifstablesinternalthin%
    \stablesinternalwidth=\stablesthinline%
  \else%
    \stablesinternalwidth=\stablesthickline%
  \fi%
  \tabskip=0pt%
  \stablesbaselineskip=\baselineskip%
  \stableslineskip=\lineskip%
  \stableslineskiplimit=\lineskiplimit%
  \offinterlineskip%
  \def\borderrule{\vrule width \stablesborderwidth}%
  \def\internalrule{\vrule width \stablesinternalwidth}%
  \def\thinline{\noalign{\hrule height \stablesthinline}}%
  \def\thickline{\noalign{\hrule height \stablesthickline}}%
  \def\trule{\omit\leaders\hrule height \stablesthinline\hfill}%
  \def\ttrule{\omit\leaders\hrule height \stablesthickline\hfill}%
  \def\tttrule##1{\omit\leaders\hrule height ##1\hfill}%
  \def\stablesel{&\omit\global\stablesmode=0%
    \global\advance\stableslines by 1\borderrule\hfil\cr}%
  \def\el{\stablesel&}%
  \def\elt{\stablesel\thinline&}%
  \def\eltt{\stablesel\thickline&}%
  \def\elttt##1{\stablesel\noalign{\hrule height ##1}&}%
  \def\elspec{&\omit\hfil\borderrule\cr\omit\borderrule&%
              \ifstablemode%
              \else%
                \errhelp=\stablelinehelp%
                \errmessage{Special ruling will not display properly}%
              \fi}%
  \def\stmultispan##1{\mscount=##1 \loop\ifnum\mscount>3 \stspan\repeat}%
  \def\stspan{\span\omit \advance\mscount by -1}%
  \def\multicolumn##1{\omit\multiply\stablestemp by ##1%
     \stmultispan{\stablestemp}%
     \advance\stablesmode by ##1%
     \advance\stablesmode by -1%
     \stablestemp=3}%
  \def\multirow##1{\stablesdummyc=##1\parindent=0pt\setbox0\hbox\bgroup%
    \aftergroup\emultirow\let\temp=}
  \def\emultirow{\setbox1\vbox to\stablesdummyc\stablestrutsize%
    {\hsize\wd0\vfil\box0\vfil}%
    \ht1=\ht\stablestrutbox%
    \dp1=\dp\stablestrutbox%
    \box1}%
  \def\stpar##1{\vtop\bgroup\hsize ##1%
     \baselineskip=\stablesbaselineskip%
     \lineskip=\stableslineskip%
     \lineskiplimit=\stableslineskiplimit\bgroup\aftergroup\estpar\let\temp=}%
  \def\estpar{\vskip 6pt\egroup}%
  \def\stparrow##1##2{\stablesdummy=##2%
     \setbox0=\vtop to ##1\stablestrutsize\bgroup%
     \hsize\stablesdummy%
     \baselineskip=\stablesbaselineskip%
     \lineskip=\stableslineskip%
     \lineskiplimit=\stableslineskiplimit%
     \bgroup\vfil\aftergroup\estparrow%
     \let\temp=}%
  \def\estparrow{\vfil\egroup%
     \ht0=\ht\stablestrutbox%
     \dp0=\dp\stablestrutbox%
     \wd0=\stablesdummy%
     \box0}%
  \def|{\global\advance\stablesmode by 1&&&}%
  \def\|{\global\advance\stablesmode by 1&\omit\vrule width 0pt%
         \hfil&&}%
  \def\vt{\global\advance\stablesmode by 1&\omit\vrule width \stablesthinline%
          \hfil&&}%
  \def\vtt{\global\advance\stablesmode by 1&\omit\vrule width
\stablesthickline%
          \hfil&&}%
  \def\vttt##1{\global\advance\stablesmode by 1&\omit\vrule width ##1%
          \hfil&&}%
  \def\vtr{\global\advance\stablesmode by 1&\omit\hfil\vrule width%
           \stablesthinline&&}%
  \def\vttr{\global\advance\stablesmode by 1&\omit\hfil\vrule width%
            \stablesthickline&&}%
  \def\vtttr##1{\global\advance\stablesmode by 1&\omit\hfil\vrule width ##1&&}%
  \stableslines=0%
  \stablesomitfalse}
\def\stablesdef{\bgroup\stablestrut\borderrule##\tabskip=0pt plus 1fil%
  &\stablesleft##\stablesright%
  &##\ifstablesright\hfill\fi\internalrule\ifstablesright\else\hfill\fi%
  \tabskip 0pt&&##\hfil\tabskip=0pt plus 1fil%
  &\stablesleft##\stablesright%
  &##\ifstablesright\hfill\fi\internalrule\ifstablesright\else\hfill\fi%
  \tabskip=0pt\cr%
  \ifstablesborderthin%
    \thinline%
  \else%
    \thickline%
  \fi&%
}%
\def\endtable{\advance\stableslines by 1\advance\stablesmode by 1%
   \message{- Rows: \number\stableslines, Columns:  \number\stablesmode>}%
   \stablesel%
   \ifstablesborderthin%
     \thinline%
   \else%
     \thickline%
   \fi%
   \egroup\stablesend%
\endgroup%
\global\stablesinfalse}
%
% end of STABLES.TEX

\overfullrule=0pt
\abovedisplayskip=12pt plus 3pt minus 3pt
\belowdisplayskip=12pt plus 3pt minus 3pt
%macros
%

%
\font\zfont = cmss10 %scaled \magstep1
\font\litfont = cmr6

\def\bigone{\hbox{1\kern -.23em {\rm l}}}
\def\ZZ{\hbox{\zfont Z\kern-.4emZ}}
\def\half{{\litfont {1 \over 2}}}

\def\mofl{(-1)^{F_L}}
\def\mofr{(-1)^{F_R}}
\def\moflfr{(-1)^{F_L+ F_R}}

% References

\def\np{Nucl. Phys.}
\def\pl{Phys. Lett.}

\def\mpl{Mod. Phys. Lett.}

\def\pr{Phys. Rev.}

\def\WITTENDYN{E. Witten, hep-th/9503124, Nucl. Phys. {\bf B443} 
(1995) 85.}
\def\VAFAF{C. Vafa, {\it ``Evidence for F-Theory''} hep-th/9602022.}
\def\DHVW{L. Dixon, J. Harvey, C. Vafa and E. Witten, \np\ {\bf B261}
(1985) 678, \np\ {\bf B274} (1986) 285.}
\def\HORAVA{P. Horava, Nucl. Phys. {\bf B327} (1989) 461.}
\def\HORETC{A. Sagnotti, {\it ``Open Strings and Their Symmetry Groups''},
in {\it Non-perturbative Quantum Field Theory}, Cargese 1987, eds. G.
Mack et. al. (Pergamon Press 1988)\semi
P. Horava, Phys. Lett. {\bf B231} (1989) 251\semi
J. Dai, R.G. Leigh and J. Polchinski, Mod. Phys. Lett. {\bf A4} (1989)
2073.}
\def\HORWIT{P. Horava and E. Witten, hep-th/9510209, \np\ {\bf B460}
(1996) 506.}
\def\SENORB{A. Sen, hep-th/9603113, \mpl\ {\bf A11}(1996) 1339.} 
\def\DMONE{K. Dasgupta and S. Mukhi, hep-th/9512196, \np\ {\bf B465}
(1996) 399.} 
\def\WITFIVE{E. Witten, hep-th/9512219, \np\ {\bf B463} (1996) 383.}
\def\SENORBM{A. Sen, hep-th/9602010, \pr\ {\bf D53} (1996) 6725.}
\def\KUMARRAY{A. Kumar and K. Ray, hep-th/9602144, \pr\ {\bf D}, to
appear.}
\def\KUMARRAYTWO{A. Kumar, K. Ray, {\it ``Compactification of M-Theory
to Two Dimensions''}, hep-th/9604164.}
\def\ACHARYAONE{B.S. Acharya, {\it ``N=1 M-Theory Heterotic Duality 
in Three-Dimensions and Joyce Manifolds''}, hep-th/9604133.}
\def\ACHARYATWO{B.S. Acharya, {\it ``M-Theory Compactification and
Two Brane/Five Brane Duality''}, hep-th/9605047.} 
\def\MVTWO{D. Morrison and C. Vafa, {\it ``Compactifications 
of F-Theory on Calabi-Yau Threefolds (II)''}, hep-th/9603161.}
\def\SENORBF{A. Sen, {\it ``F-Theory and Orientifolds''}, 
hep-th/9605150.}
\def\DMTWO{K. Dasgupta and S. Mukhi, {\it ``F-Theory at Constant
Coupling''}, hep-th/9606044.}
\def\GIMPOL{E. Gimon and J. Polchinski, hep-th/9601038, \pr\ {\bf D}, 
to appear.}
\def\DPONE{A. Dabholkar and J. Park, {\it ``An Orientifold 
of Type IIB Theory on K3''}, hep-th/9602030.}
\def\GJONE{E. Gimon and C. Johnson, {\it ``K3 Orientifolds''},
hep-th/9604129.}
\def\DPTWO{A. Dabholkar and J. Park, {\it ``Strings on Orientifolds''},
hep-th/9604178.}
\def\BERKL{M. Berkooz and R. Leigh, {\it ``A D=4 N=1 Orbifold 
of Type I Strings''}, hep-th/9605049.}
\def\BLPSSW{M. Berkooz, R. Leigh, J. Polchinski, J.H. Schwarz, N.
Seiberg and E. Witten, 
{\it ``Anomalies, Dualities, and Topology of D=6 N=1 Superstring
Vacua''}, hep-th/9605184.}
\def\POLKTHREE{J. Polchinski, {\it ``Tensors From K3 Orientifolds''}, 
hep-th/9606165.}
\def\VOISIN{C. Voisin, in {\it Journ\'ees de G\'eometrie
Alg\'ebrique d'Orsay}, Ast\'erisque No. {\bf 218} (1993) 273.}
\def\BORCEA{C. Borcea, {\it ``K3 Surfaces with Involution and Mirror
Pairs of Calabi-Yau Manifolds''}, to appear in {\it Essays on Mirror
Manifolds II}. }
\def\BIKMSV{M. Bershadsky, K. Intriligator, S. Kachru, D. Morrison, V.
Sadov and C. Vafa,
{\it ``Geometric Singularities and Enhanced Gauge Symmetries''}, 
hep-th/9605200.}
\def\FHSV{S. Ferrara, J. Harvey, A. Strominger and C. Vafa, 
hep-th/9505162, \pl\ {\bf B361} (1995) 59.}
\def\DMW{M. Duff, R. Minasian and E. Witten,
{\it ``Evidence For Heterotic/Heterotic Duality''}, hep-th/9601036.}
\def\MVONE{D. Morrison and C. Vafa, {\it ``Compactifications 
of F-Theory on Calabi-Yau Threefolds 1''}, hep-th/9602114.}
\def\SCHWARZSEN{J.H. Schwarz and A. Sen, hep-th/9507027, 
\pl\ {\bf B357} (1995) 323.}
\def\CHAUDHURI{S. Chaudhuri and D. Lowe, hep-th/9508144, \np\ {\bf B459} 
(1995) 113.}
\def\VAFADISC{C. Vafa, \np\ {\bf B273} (1986) 592.}
\def\FIQDISC{A. Font, L.E. Ibanez and F. Quevedo,  
\pl\ {B217} (1989) 272.}
\def\VWDISC{C. Vafa and E. Witten, hep-th/9409188,
J. Geom. Phys. {\bf 15} (1995) 189.}
\def\AMDISC{P. Aspinwall and D. Morrison (appendix by M. Gross),
hep-th/9503208, Comm. Math. Phys. {\bf 178} (1996) 115.}
\def\NIKULIN{V. Nikulin, in {\it Proceedings of the International
Congress of Mathematicians, Berkeley}, (1986) 654.}
\def\SVW{S. Sethi, C. Vafa and E. Witten, {\it ``Constraints on 
Low-Dimensional String Compactifications''}, hep-th/9606122.}
\def\SENORBDUAL{A. Sen, {\it ``Duality and Orbifolds''},
hep-th/9604070.}
\def\JOYCE{D. Joyce, {\it ``Compact Riemannian Manifolds with Holonomy
Spin(7)''}, to appear in Inv. Math.}
\def\SEIWIT{N. Seiberg and E. Witten, {\it ``Comments on String 
Dynamics in Six-Dimensions''}, hep-th/9603003.}
\def\WITPHASEMF{E. Witten,
{\it ``Phase Transitions in M-Theory and F-Theory''},
hep-th/9603150.}
\def\WITNONPERT{E. Witten,
{\it ``Nonperturbative Superpotentials in String Theory''},
hep-th/9604030.}
\def\BRUNSCHIM{I. Brunner and R. Schimmrigk, {\it ``F-Theory 
on Calabi-Yau Fourfolds''}, hep-th/9606148.}
\def\GJTWO{E. Gimon and C. Johnson, {\it ``Multiple Realizations
of N=1 Vacua in Six-Dimensions''},
hep-th/9606176.}
\def\BLUMZAF{J. Blum and A. Zaffaroni, {\it ``An Orientifold from
F-Theory''}, hep-th/9607019.}
\def\DPTHREE{A. Dabholkar and J. Park, {\it ``A Note on Orientifolds
and F-Theory''}, hep-th/9607041.}

{\nopagenumbers
\Title{\vtop{\hbox{hep-th/9607057}
\hbox{PUPT-1636}
\hbox{TIFR/TH/96-37}
\hbox{July 1996}}}
{\advance\baselineskip by 6pt
\vtop{\centerline{Orbifold and Orientifold Compactifications of}
\centerline{F-Theory and M-Theory to Six and Four Dimensions}}}
\centerline{Rajesh Gopakumar\foot{E-mail: gopakumr@phoenix.princeton.edu}}
\vskip 2pt
\centerline{\it Department of Physics, Princeton University,}
\centerline{\it Princeton, NJ 08544, U.S.A.}
\vskip 8pt
\centerline{and} 
\vskip 8pt
\centerline{Sunil Mukhi\foot{E-mail: mukhi@theory.tifr.res.in}}
\vskip 2pt
\centerline{\it Tata Institute of Fundamental Research,}
\centerline{\it Homi Bhabha Rd, Mumbai 400 005, India}
\ \smallskip
\centerline{ABSTRACT}

{\advance\baselineskip by -2pt We study orbifold compactifications of
F-theory which lead to $N=1$ supersymmetry in 6 and 4 spacetime
dimensions. These are dual to specific orientifolds of M-theory, and
in many cases to orientifolds of type IIB string theory. The
equivalences are demonstrated by mapping the orbifolding
transformations in the F, M and string theories to each other using
dualities. We observe that M and F-theory appear to possess a property
similar to discrete torsion in string theory. This is related to an
ambiguity recently noted by Polchinski in the orientifold projection
for 6-dimensional models. The 4-dimensional compactifications exhibit
similar features, from which we predict the existence of certain new
orientifolds of type IIB.  Some orbifolds with higher supersymmetry
are also examined.

}
\vfill\eject} 
\ftno=0
\newsec{Introduction} 
Many physical and mathematical phenomena that have been regarded as
unique to string theory turn out to have deeper explanations in terms
of M-theory\ref\wittendyn{\WITTENDYN}\ and F-theory\ref\vafaf{\VAFAF}.

Compactifications of these new theories are in general difficult to
describe, since we presently lack a formulation of these theories with
the same power as perturbative string theory. However, in string
theory it has often been the case that such compactifications are more
easily studied for orbifolds\ref\dhvw{\DHVW}\ (and their
generalizations,
orientifolds\ref\horava{\HORAVA}\ref\horetc{\HORETC}). Recently it has
become evident that compactifications of M and F-theory on these
spaces can also be discussed in a surprising amount of detail.

The first example of this sort was the orientifold of M-theory on
$S^1/Z_2$, shown by Horava and Witten\ref\horwit{\HORWIT}\ to be dual
to the $E_8 \times E_8$ heterotic string in 10 dimensions.  This
example turns out not to be sufficiently generic; for example, this is
a case where orbifolding does not commute with
duality\ref\senorb{\SENORB}. The first generic examples were the
orientifolds of M-theory on $T^5/Z_2$ down to 6 dimensions, and on
$T^9/Z_2$ down to 2 dimensions, described in
Ref.\ref\dmone{\DMONE}. These were argued to be dual to the type IIB
string compactified on K3 and $T^8/Z_2$ respectively. The $T^5/Z_2$
case, independently studied in Ref.\ref\witfive{\WITFIVE}, has another
interesting property: the vacuum contains 5-branes of M-theory, which
carry the requisite matter multiplets to cancel the anomaly in 6d.
Subsequently, a number of M-theory orientifolds\ref\senorbm{\SENORBM}%
\ref\kumarray{\KUMARRAY}\senorb\ref\kumarraytwo{\KUMARRAYTWO}\ and
orbifolds\ref\acharyaone{\ACHARYAONE}\ref\acharyatwo{\ACHARYATWO}\
have been investigated. (We distinguish between orbifold and
orientifold compactifications of M-theory by the fact that the latter
have an orientiation-reversing action on the compactification torus
and reverse the M-theory 3-form.)

For F-theory, orbifold limits of certain Calabi-Yau 3-fold
compactifications were discussed in Ref.\ref\mvtwo{\MVTWO}. Orbifold
limits of the original K3 compactification have been recently
investigated in some detail in
Refs.\ref\senorbf{\SENORBF}\ref\dmtwo{\DMTWO}. Most of the examples
discussed in the above references dealt with compactification down to
8 or 6 spacetime dimensions.

Studying orbifold/orientifold compactifications of F and M theory is
likely to be particularly useful because we still lack a satisfactory
fundamental formulation of these theories. Such a formulation would
also presumably be a fundamental non-perturbative formulation of
string theory. We may hope to gain some insight into this problem by
using flat but singular compactifying spaces.

A distinct but closely related programme pursued in recent times has
been the study of orientifolds of string
theory\ref\gimpol{\GIMPOL}\ref\dpone{\DPONE}\ref\gjone{\GJONE}%
\ref\dptwo{\DPTWO}\ref\berkl{\BERKL}\ref\blpssw{\BLPSSW}%
\ref\polkthree{\POLKTHREE}.
These have the advantage of being free
conformal field theory realisations of non-perturbative vacua.
Some inter-relationships among 
orientifolds of string theory, and orbifolds/orientifolds of M and F
theories, were investigated in Refs.\senorb\senorbf\dmtwo.

In what follows, we study a class of F-theory compactifications on
Calabi-Yau threefolds and fourfolds down to 6 and 4 spacetime
dimensions. These give rise in each case to vacua with the minimum
supersymmetry compatible with the number of spacetime dimensions. The
special property of the manifolds we study is the fact that they admit
orbifold limits in their moduli space.  It is then possible to
establish an equivalence to orientifolds of M-theory of the generic
form $T^5/(Z_2)^2$ and $T^7/(Z_2)^3$ respectively.  These dualities
can also be extended to produce type IIB orientifold duals, some of
which have been independently investigated in the various recent works
cited above. We will see that while the map between F-theory and IIB
orientifolds is perturbative, that to M-theory is not. This allows us
to view these vacua often from complementary viewpoints.  This is
potentially useful in understanding interesting objects like
tensionless strings which play a role in phase transitions.  One
example of this utility will be in examining an ambiguity in IIB
orientifolds that we will describe later.  Our mapping also leads us
to believe that many M-theory orientifold vacua may be realised in
F-theory.

In the four-dimensional case, we actually study a much larger class of
non-toroidal M-theory orientifolds, of the form $CY_3\times S^1/Z_2$ ,
$(K3\times T^2 \times S^1/Z_2) /(Z_2)$ etc., for which again the
equivalences to F theory and IIB orientifolds are demonstrated. These
compactifications all have N=1 spacetime supersymmetry in 4d. Given
the potential importance of four-dimensional N=1 supersymmetry to
present-day particle physics, it is clearly desirable to examine
properties of the most solvable cases in some detail. We believe the
cases at hand fall into this category.

We find that for orbifolding groups larger than $Z_2$, the projection
of twisted-sector states of one element onto the subspace invariant
under the other elements sometimes presents ambiguities. In particular
for the group $Z_2\times Z_2$ we find a model with a twofold
ambiguity, leading to two distinct vacua. The phenomenon has a close
operational resemblance to discrete torsion in string theory, and we
develop the analogy in some detail. For $Z_2\times Z_2\times Z_2$
models there is apparently a fourfold ambiguity. We relate the twofold
ambiguity to a phenomenon discussed recently by Polchinski for string
theory on orientifolds to six dimensions. 
The fourfold ambiguity
appears to be a combination of the above phenomenon and conventional
discrete torsion, and seems to indicate that the two must enter on
an equal footing in F-theory or M-theory.

Most, but not all, of the compactifications of F-theory to 4
dimensions that we define pass the consistency test of
Ref.\ref\svw{\SVW}, according to which the Euler characteristic $\chi$
should be a positive integral multiple of 24, so that tadpoles can be
cancelled by 3-branes in the vacuum. We will find one model with
$\chi=0$, for which no such 3-branes are required.

This paper is organized as follows. In Sections 2 and 3 we define and
analyse orbifold/orientifold compactifications of F and M theory to 6
dimensions, and demonstrate their equivalence to each other and to
orientifolds of type IIB. The ambiguity referred to above is discussed
in Section 3. In Sections 4 and 5 we discuss analogous
compactifications to 4 dimensions. We make a number of predictions
which could be verified by further investigations of type IIB
orientifolds. In section 6 we discuss some orientifolds and orbifolds
of M and F theory which lead to $N>1$ supersymmetry in 4
dimensions.  
Finally, in Section 7 we present the conclusions and
describe some open questions for the future.

\newsec{Compactifications to six dimensions}

We start by describing a set of three Calabi-Yau complex threefolds,
all of which contain a point in their moduli space where they can be
realised as orbifolds of the form $T^6/(Z_2\times Z_2)$. All of them
are actually members of the Voisin-Borcea family of Calabi-Yau
3-folds\ref\voisin{\VOISIN}\ref\borcea{\BORCEA}, and have already been
studied in connection with string theory and F-theory
compactifications in Ref.\mvtwo.

In what follows we shall study F-theory on these manifolds. This may
be regarded as IIB compactified on the base $T^4/(Z_2\times Z_2)$ with
the $T^2$ fibre typically degenerating to $T^2/Z_2$ on fixed
points/curves on this base.  Let us briefly recapitulate how to obtain
the spectrum for F-theory on Calabi-Yau 3-folds\mvtwo.  Upon further
compactification on $T^2$, the theory is on the same moduli space as
IIA on the 3-fold. This immediately tells us that the number of
neutral hypermultiplets is
\eqn\hyp{h^0=H^{2,1}(X)+1.}
The tensor multiplets are determined by the Kahler deformations that
preserve the elliptic fibration. Moreover since vectors and tensors of
N=1 in 6d on compactification give vectors of N=2 in 4d, we have
\eqn\vect{T=H^{1,1}(B)-1 ,\ \  r(V)= H^{1,1}(X)-H^{1,1}(B)-1}
Here $X$ and $B$ refer to the 3-fold and the base of the fibration
respectively, $T$ is the number of N=1 tensor multiplets and $r(V)$ is
the rank of the non-abelian gauge group. The gauge group itself is
determined by the singularity type of the degenerating fibre. There
are subtleties in this dictionary\ref\bikmsv{\BIKMSV}, but we will
choose to work in the region of moduli space where the naive gauge
group is realised.

In what follows, we label the spacetime coordinates $x^1,x^2,\ldots$
where $x^1$ is the time. The orbifolds can be defined as follows: each
of the three non-trivial elements of order 2 in the orbifolding group
$Z_2 \times Z_2$ reverses the sign of 4 of the 6 torus coordinates.
Labelling these three elements $\alpha$, $\beta$ and
$\alpha\beta$, the orbifold action is given by specifying which
of the 6 coordinates $x^7,x^8,\ldots,x^{12}$ is inverted, and whether
the coordinate also gets a shift by a $1/2$-unit. The ``coordinates''
$x^{11},x^{12}$ will be taken to be those of the fibre $T^2$. There
are three inequivalent choices which give us in turn the following
models, labelled by their Hodge numbers $(H^{1,1}, H^{2,1})$:
\bigskip
\centerline{Model A: (11,11)}\nobreak
\medskip
\begintable
|$x^{12}$|$x^{11}$|$x^{10}$|$x^9$|$x^8$|$x^7$\elt
$\alpha$|$+$|$+$|$-$|$-$|$-$|$-$\elt
$\beta$|$-$|$-$|$+,\half$|$+$|$-,\half$|$-$\elt
$\alpha\beta$|$-$|$-$|$-,\half$|$-$|$+,\half$|$+$
\endtable

This manifold has also featured in Type II
compactifications\ref\fhsv{\FHSV}. The base is the so-called Enriques
surface. Note that it has no fixed points where the fibre
degenerates. Hence one expects no gauge fields at all. The spectrum
thus consists of 9 tensor and 12 hypermultiplets. The same spectrum
also appears in type IIB on a $T^4/Z_4$ orientifold\gjone\dptwo.

\bigskip
\centerline{Model B: (19,19)}\nobreak
\medskip
\begintable
|$x^{12}$|$x^{11}$|$x^{10}$|$x^9$|$x^8$|$x^7$\elt
$\alpha$|$+$|$+$|$-$|$-$|$-$|$-$\elt
$\beta$|$-$|$-$|$+,\half$|$+$|$-$|$-$\elt
$\alpha\beta$|$-$|$-$|$-,\half$|$-$|$+$|$+$
\endtable

Here we see that the fibre degenerates to $T^2/Z_2$ at the four fixed
points of $\alpha\beta$. The monodromy around these points is
precisely of the form studied in\senorbf. We know that it is a $D_4$
singularity and results in an $SO(8)$ enhanced symmetry coming from
four coincident 7-branes located at each fixed point (and wrapped
around $x^{7},x^{8}$). The four fixed points are, however, permuted
pairwise by the action of $\beta$, and thus we only get an $SO(8)^2$
enhanced symmetry. By moving the 7-branes off the fixed points, we
generically break the symmetry to $U(1)^8$. Thus at generic points in
its moduli space, this model has 9 tensor and 20 neutral
hypermultiplets. The Higgsing at the orbifold point was accomplished
by 2 hypermultiplets in the adjoint of $SO(8)$. The gauge symmetries
can be enhanced at other special points of moduli space as well, with
the addition of charged hypermultiplets. This spectrum coincides with
that of an M-theory orbifold\senorbm\ and a IIB orientifold\dpone.
Later we will demonstrate the equivalence of all three realisations.

\bigskip
\centerline{Model C: (51,3)}\nobreak
\medskip
\begintable
|$x^{12}$|$x^{11}$|$x^{10}$|$x^9$|$x^8$|$x^7$\elt
$\alpha$|$+$|$+$|$-$|$-$|$-$|$-$\elt
$\beta$|$-$|$-$|$+$|$+$|$-$|$-$\elt
$\alpha\beta$|$-$|$-$|$-$|$-$|$+$|$+$
\endtable

This last model shows many intriguing features, as we will see.  First
note that the the base $T^4/(Z_2 \times Z_2)$ is actually $T^2/Z_2
\times T^2/Z_2$ with the $Z_2$'s generated by $\beta$ and
$\gamma$. The fibre degenerates over $4+4$ curves of the base in the
manner of\senorbf. There are also $4 \times 4$ fixed points on the
base, which can be blown up and thus contribute 16 to the total
$H^{1,1}(B)=18$.  Hence we have an anomaly-free spectrum consisting of
$SO(8)^8$ vector multiplets, 17 tensor multiplets and 4 neutral
hypermultiplets. The absence of charged hypermultiplets implies that
the gauge group is generically non-abelian \mvtwo\ and can even be
further enhanced.

The fact that the base looks like a singular limit of $P^1\times P^1$
suggests a connection with another Calabi-Yau, the (3,243), which
admits an elliptic fibration over $P^1\times P^1$.  This latter model
is dual to the $E_8\times E_8$ heterotic string on $K3$ with symmetric
instanton embedding\ref\dmw{\DMW}\ref\mvone{\MVONE}. This connection
will be examined in more detail below.

As we noted above, for the (19,19), there is a five-dimensional
M-theory orientifold\senorbm, and also a type IIB orientifold\dpone\
with the same spectrum. M-theory duals to the other two cases have not
so far been identified. We now show that there are precisely three
M-theory orientifolds of the form $T^5/(Z_2\times Z_2)$ which give N=1
supersymmetry in 6d, and can be put in a natural correspondence with
the Calabi-Yau orbifolds A, B and C of F-theory described above, and
that M-theory compactified on each of them can be mapped in a precise
way to F-theory on the corresponding Calabi-Yau.

To this end, first note that in the (11,11), only one of the three
nontrivial elements of the orbifold group has fixed loci, while in the
(19,19), two elements have fixed loci and in the (51,3), all three
have fixed loci. For an M-theory orbifold to have N=1 supersymmetry
one also needs two $Z_2$ actions, each of which reduces the
supersymmetry by half. It is easy to see that this can only happen if
the number of coordinate inversions is 0 or 1 mod 4. This forces the
three nontrivial elements of $Z_2\times Z_2$ to consist of 5, 4 and 1
inversion in turn. It remains to incorporate shifts in appropriate
ways so that one is led to three models where one, two and three
elements act with fixed points. Thus we are led to consider M-theory
on the three orbifolds (more properly labelled as orientifolds), of
the form $T^5/Z_2\times Z_2$ defined below. The spectra of these
three models will also be worked out in the following. The untwisted
sector is common to all three models and is easily shown to consist of
one tensor and four neutral hypermultiplets. Thus we need only examine
the twisted sector separately for each case.
\bigskip
\centerline{Model A$'$: $T^5/(Z_2\times Z_2)$}\nobreak
\medskip
\begintable
|$x^{11}$|$x^{10}$|$x^9$|$x^8$|$x^7$\elt
$\alpha$|$-$|$-$|$-$|$-$|$-$\elt
$\beta$|$+,\half$|$-$|$-$|$-,\half$|$-$\elt
$\alpha\beta$|$-,\half$|$+$|$+$|$+,\half$|$+$
\endtable

The element $\alpha$ is the only element that acts with fixed points.
It's action on $T^5$ leads to the $T^5/Z_2$ orbifold of
M-theory\dmone\witfive.The element $\beta$ is seen to be a shift in
the 11th direction and an action on the remaining $T^4$, which is
identical to the involution on $T^4/Z_2$ considered
in\ref\schwarzsen{\SCHWARZSEN}\ref\chaudhuri{\CHAUDHURI}. The twisted
sector of $\alpha$ has been argued to consist of 16 M-theory
5-branes\witfive, each carrying a multiplet that in $N=1$
supersymmetry is the sum of a tensor and a hypermultiplet.  In the
present case, this twisted-sector multiplet needs to be projected onto
the sector invariant under $\beta$ and $\gamma$. Since the
twisted-sector states arise from 5-branes, it is natural to suppose
that the branes are to be located at 16 of the fixed points of the
orientifold, and the action of $\beta$ and $\gamma$ on them is just
the respective geometrical action. Because of the shifts, this means
that the 16 5-branes get permuted amongst each other, with the result
that half of them are odd and get projected out. This leaves 8 tensor
and 8 hypermultiplets, which when combined with the untwisted sector
give precisely the same spectrum as for F-theory on model A.

\bigskip
\centerline{Model B$'$: $(K3\times S^1)/Z_2$ }\nobreak
\medskip
\begintable
|$x^{11}$|$x^{10}$|$x^9$|$x^8$|$x^7$\elt
$\alpha$|$-$|$-$|$-$|$-$|$-$\elt
$\beta$|$+$|$-,\half$|$-$|$-$|$-$\elt
$\alpha\beta$|$-$|$+,\half$|$+$|$+$|$+$
\endtable

For model B$'$, there are two elements acting with fixed points.  The
element $\alpha$ is the same as before but $\beta$ differs from that
of A$'$ in not having the shift in $x^{11}$. This makes it, in fact,
the orbifold limit of $(K3\times S^1)/Z_2$ considered in
Ref.\senorbm, but we go through the analysis here again to show
how it relates to the other cases. For $\alpha$, the twisted sector
just gives the 16 5-branes as before, which are again projected to
half their number. But now the element $\beta$ also has fixed
points. This time the twisted sector is that of IIA on $T^4/Z_2$ ,
which in six-dimensional $N=1$ language has 16 vector and 16
hypermultiplets. Generically the gauge group is thus $U(1)^{16}$, which
can be enhanced at special points upto $E_8\times E_8$. These have to
be projected onto the sector invariant under the remaining
generators. Once again, the action of $\alpha\beta$ permutes the
vector multiplets pairwise, and so half of them will be invariant,
leading to 8 vectors and 8 hypers.  Putting everything together, one
finds at generic points 9 tensors, 8 vectors and 20 hypermultiplets,
the same as for F-theory on model B.
\bigskip
\centerline{Model C$'$: $K3\times S^1/Z_2$ }\nobreak
\medskip
\begintable
|$x^{11}$|$x^{10}$|$x^9$|$x^8$|$x^7$\elt
$\alpha$|$-$|$-$|$-$|$-$|$-$\elt
$\beta$|$+$|$-$|$-$|$-$|$-$\elt
$\alpha\beta$|$-$|$+$|$+$|$+$|$+$
\endtable

Finally, for model C$'$, the situation is more complicated.  All the
elements act with fixed points. In fact the action simply corresponds
to $T^4/Z_2\times S^1/Z_2$. In this case, the 16 5-branes are not
permuted by the other symmetries.  Rather, if they are located at
their fixed points then the other actions leave them fixed. However,
it can be argued that near any of the fixed points, four of the five
transverse collective coordinates of the 5-brane get reversed by the
action of the element $\beta$. As a result, the hypermultiplet degrees
of freedom of each 5-brane get projected out and we are left with 16
tensor multiplets. This leaves the vector multiplets, which correspond
to a rank 32 gauge group before projection (rank 16 from the twisted
sector of $\beta$, which we have studied above, and another rank 16
from the twisted sector of $\alpha\beta$ which corresponds to M-theory
on $S^1/Z_2$ (further compactified on $T^4$) and hence produces again
the gauge sector of the heterotic string). This time it is easy to see
that an abelian $U(1)^{16}$ gauge group is inconsistent with anomaly
cancellation. Indeed, if $v$ and $h$ are the number of vector and
hypermultiplets coming from these twisted sectors after projection,
and $V$, $H$ and $T$ are the total numbers of vector, hyper and tensor
multiplets in the model, then the anomaly cancellation condition
\eqn\anomaly{H - V = 273 - 29T}
requires that we have
\eqn\anom{
h - v = -224}
This shows that we must have at least 224 vector multiplets, and
moreover any additional vector multiplets will be accompanied by
charged hypers. Thus the simplest solution is $v=224$ and $h=0$. This
in turn can arise symmetrically only if we are at a point of enhanced
symmetry with $S0(8)^8$ gauge group, and the projection eliminates all
the hypermultiplets that come along with the gauge multiplets. So we
conclude that this orientifold gives a spectrum of 17 tensor, 224
vector and 4 hypermultiplets, precisely that of F-theory on model C at
generic points.

Although the arguments in the last case seem less compelling than in
the previous two, one can point out some intriguing properties in the
M-theory picture. The two sets of vector multiplets that arise from
the twisted sectors of $\beta$ and $\alpha\beta$ respectively, are
believed to arise in M-theory by the wrapping respectively of a
5-brane around directions 7,8,9,10, and of a 2-brane around direction
11. If there is an electric-magnetic 2-brane-5-brane duality in
M-theory, then it means that one $S0(8)^4$ factor of the gauge group
is electric-magnetic dual to the other, which might conceivably be
checked once such a duality is better understood. In a heterotic
picture this translates into perturbative and non-perturbative
contributions of $SO(8)^4$ to the gauge group\dmw.

In studying the above model, we used the fact that the natural action
of the element $\beta$ on the 5-brane multiplets was to reverse the
sign of four of the five transverse brane coordinates. One may argue
that it is equally natural to concentrate on $\gamma$ instead of
$\beta$, in which case the natural action would be to reverse one of
the five transverse brane coordinates. (In this case the element
$\beta$ would have the opposite action to its natural one.) This will
be consistent with supersymmetry only if the coordinate projected out
is part of an N=1 tensor multiplet, and if the accompanying tensor
field also gets projected out. The result is then that we retain 16
hypermultiplets from the 5-brane sector, and project out all 16 tensor
multiplets.

The corresponding model now has only one tensor multiplet (from the
untwisted sector), and at least 20 hypers, of which 4 come from the
untwisted sector and another 16 from the branes. The remaining states
come from the twisted sectors of $\beta$ and $\gamma$. If we again
denote by $h$ and $v$ the number of hyper and vector multiplets coming
from these twisted sectors after projection, anomaly cancellation this
time requires
\eqn\anomtwo{
h - v = 224}
The simplest solution of this arises by assuming that there are no
vector multiplets at a generic point in moduli space, in which case we
have $h=224$. The model therefore has a total of one tensor, no
vectors and 244 neutral hypermultiplets at a generic point. This
coincides with the generic spectrum of F-theory on the well-known
(3,243) Calabi-Yau , dual to the (12,12) embedding of heterotic string
on K3.  In fact, precisely $K3\times S^1/Z_2$ was considered in
Ref.\dmw\ to argue for heterotic-heterotic duality in this case.

It appears, therefore, that potentially there are two distinct
versions of model C$'$. If this is the case, there must be two
distinct ways of smoothing the orbifold C in F-theory. Although we do
not have a complete understanding of this situation, it appears to
strongly resemble the phenomenon of discrete
torsion\ref\vafadisc{\VAFADISC}\ in Calabi-Yau compactifications of
type II string theory. There, it is
known \ref\fiqdisc{\FIQDISC}
\ref\vwdisc{\VWDISC}\ref\amdisc{\AMDISC}\ that there are two
distinct theories associated to the orbifold that we have defined as
model C above. The standard blowup, corresponding to deforming the
K\"ahler structure, leads to the (51,3) Calabi-Yau. This is the theory
that one would have discovered following the usual rules for
constructing orbifold compactifications in string theory, since the
twisted sector modes that string theory produces are exactly the
marginal deformations of K\"ahler structure required to smooth the
orbifold to the (51,3) Calabi-Yau.

However, one can instead consistently require that the elements
$\beta$ and $\gamma$ of the orbifolding group act with an extra minus
sign on twisted-sector states of $\alpha$, and similarly for the other
two twisted sectors. This choice was interpreted in 
Refs.\fiqdisc \vwdisc\ as
arising from the presence of ``discrete torsion''. (A more detailed
analysis of this issue in terms of Brauer groups appeared in
Ref.\amdisc.) In this case, one finds that the twisted sector states
correspond to deformations of the complex, rather than K\"ahler,
structure, and the orbifold gets deformed to a (singular) Calabi-Yau
manifold which still retains 64 nodes, with Hodge numbers (3,51).

It should be evident that the two versions of model C$'$ that we have
discovered (which we denote henceforth by C$'$(1) and C$'$(2)) have a
lot in common with the discrete torsion problem discussed above: in
the twisted sector of $\alpha$, model C$'$(1) keeps 16 tensor
multiplets and projects out 16 hypers, while model C$'$(2) does the
opposite. Similarly, in the twisted sectors for $\beta$ and $\gamma$,
model C$'$(1) keeps a total of 224 vector multiplets, which give rise
to the gauge group $SO(8)^8$, while model C$'$(2) keeps instead 224
hypermultiplets, which are of course neutral as there is no gauge
group. Thus complementary sets of multiplets are projected in/out.
The somewhat surprising conclusion is that the analogue of discrete
torsion in M-theory (and, as we argue below, in F-theory) relates two
smooth Calabi-Yau's with Hodge numbers (51,3) and (3,243), rather than
relating the smooth (51,3) to its ``mirror'', the (3,51) with 64
nodes.

Let us briefly examine how the same ambiguity arises in F-theory on
model C. Here, the twisted sector for $\beta$ produces 24 F-theory
7-branes, of which 6 are located at each of the 4 fixed points of
$\beta$ on the $(x^7,x^8)$ torus, and all are wrapped around the
$(x^9,x^{10})$ torus. The twisted sector for $\alpha\beta$ is
analogous, with the roles of $(x^7,x^8)$ and $(x^9,x^{10})$
interchanged.  Finally, the twisted sector for $\alpha$ just consists
of the usual twisted-sector states of type IIB on K3, which are
well-known to correspond to 16 tensor multiplets of N=2, equivalent to
16 tensor and 16 hypermultiplets of N=1. The projection leading to
F-theory on (51,3) keeps all the N=1 tensor multiplets, while that
leading to F-theory on (3,243) keeps the hypermultiplets. We will not
discuss the 7-brane sectors here since they are known to be quite
subtle in F-theory\vafaf, but the observations above should be
sufficient to see that the discrete-torsion-like phenomenon, of
projecting in/out complementary sets of states, operates in F-theory
too. It is amusing to note a different kind of complementarity between
M- and F-theory in this model (and the other ones discussed above):
multiplets supported on branes in either description are
``fundamental'' states in the other one.

\newsec{Mappings between M-theory, F-theory and IIB orientifolds}

Having discussed three pairs of candidate M-theory-F-theory duals, we
will now demonstrate the proposed equivalence
as also the relation to some familiar and unfamiliar IIB orientifolds.
The strategy will be to map the symmetries of one theory to the other
at the orbifold point.  Since the spectra are independently known to
coincide, identifying the map between symmetries automatically implies
that orbifolding commutes with duality. (The conditions under which
orbifolding commutes with duality have been recently examined in a
variety of situations\senorb. Our case is of the type where this
is known to hold in a variety of cases.)

We first demonstrate the equivalence upon compactification to
5 dimensions. Since
F-theory on $X\times S^1$ is on the same moduli space as M-theory on
$X$, this has the advantage of reducing the problem to that between
two different M-theory compactifications. But we will, in a more
direct manner, also be able to show the equivalence in 6d by mapping
the orbifolding symmetries in F-theory to those of IIB. This will also
make transparent the relation to the IIB orientifold.

Thus, to start with, we compactify the models of Section 2 on a
circle, so that the question reduces to duality between M-theory on
the Calabi-Yau's given by models A,B and C above, and type IIA on the
corresponding orientifolds A$'$,B$'$ and C$'$. Let us focus on models
C, the (51,3), and C$'$ for definiteness and simplicity. Since we have
compactified to one lower dimension, the coordinate labellings for the
various orbifold actions will all have to be shifted down by 1 unit
with respect to those displayed in the tables.

The symmetries of M-theory can be mapped to those of type IIA as
follows: An inversion in the 11th direction is mapped to $\mofl$ where
$F_L$ is the spacetime fermion number for left-moving states on the
worldsheet\senorb. Also an inversion of an odd number of spatial
dimensions in type IIA must be accompanied by a world-sheet
orientation reversal, $\Omega$, to be a symmetry.  We can now easily
read off from table C that M-theory on the (51,3) can be written as an
orbifold of type IIA on $T^5$ modded out by $\{1, I_{6789},$
$\Omega\mofl I_{6710},$ $\Omega\mofl I_{8910}\}$. Here $I_{...}$ 
represents inversions in the relevant directions.

For future use, we list some of the relevant properties of the symbols
that we will use. The square of an inversion acts as $\pm 1$ on
spacetime fermions according to
\eqn\properties{
\eqalign{
I_{n_1\ldots n_k}^2 &= 1,\qquad k=0,1~{\rm mod}~4\cr
 &= \moflfr,\qquad k=2,3~{\rm mod}~4\cr}}
as one can easily show using $\Gamma$-matrices. Orientation reversal
does not commute with reversing the sign of left-moving fermions, and
in fact we get
\eqn\moreprop{
\eqalign{ \Omega\mofl\Omega &= \mofr\cr
(\Omega\mofl)^2 &= \moflfr\cr}}
{}From these facts one can check that all elements of the action
described above are of order two as they should be.

Some additional results that we will need state that (at least for
orbifold groups involving only $Z_2$ factors) S-duality of type IIB
interchanges $\Omega$ and $\mofl$, that under T-duality on all $k$
directions ($T_{n_1\ldots n_k}$) the inversion $I_{n_1\ldots n_k}$
goes to $(-1)^{kF_L}I_{n_1\ldots n_k}$, and that under T-duality the
orientation-reversal $\Omega$ picks up an inversion in all the
dualized directions, along with a possible factor of $\mofl$ which can
be most simply determined by requiring that all elements be of order
2.

Now, we have the following chain of equivalences:
\eqn\chain{
\eqalign{
&{\rm IIA~on}~T^5/\{1, I_{6789}, \Omega\mofl I_{6710}, \Omega\mofl 
I_{8910}\}\cr
{\buildrel T_{8}\over\longrightarrow}~ & 
{\rm IIB~on}~T^5/\{1, \mofl I_{6789}, \Omega I_{67810}, \Omega\mofl
I_{910}\}\cr
{\buildrel S\over\longrightarrow}~ & 
{\rm IIB~on}~T^5/\{1, \Omega I_{6789}, \mofl I_{67810}, \Omega\mofl 
I_{910} \}\cr
{\buildrel T_{10}\over\longrightarrow}~ & 
{\rm IIA~on}~T^5/\{1, \Omega I_{678910}, I_{67810}, \Omega
I_{9}\}\cr
{\buildrel 9\leftrightarrow 10\over\longrightarrow}~ & 
{\rm IIA~on}~T^5/\{1, \Omega I_{678910}, I_{6789}, \Omega I_{10}\}\cr} }

Comparison of the last line with table C$'$ reveals that it
corresponds precisely to type IIA on the orientifold model C$'$. The
same manipulations can be used to prove the equivalence of M-theory on
the Calabi-Yau's (11,11) (model A) and (19,19) (model B) to type IIA
on the orientifolds $T^5/(Z_2\times Z_2)$ (A$'$) and $(K3\times
S^1)/Z_2$ (B$'$) respectively, the only difference being that one has
to incorporate the relevant shifts. Since these act in directions
along which we have not T-dualized in the above chain, they go through
unaffected.

Thus, at least after compactifying to 5 dimensions, the proposed
equivalence of F-theory on orbifolds and M-theory on orientifolds can
be demonstrated, not just by matching spectra, but by directly
identifying symmetries of the two models, and orbifolding by the
corresponding symmetries on each side. This may perhaps be enough to
prove equivalence in 6 dimensions, for the following reason. In the
(11,11) case, since there are no vector fields in the 6d spectrum, the
5d equivalence which relates the vectors to vectors must come from a
6d equivalence in which the tensors are mapped to tensors. The map
between hypermultiplets is the same in 5 and 6 dimensions. Thus there
is no ambiguity in going back to 6 from 5 dimensions in this case. For
the (19,19), in 5 dimensions we have 18 vector multiplets.
Discounting the one coming from compactification and one more coming
from an untwisted-sector tensor in 6d, the remaining 16 come from 8
vectors and 8 tensors in 6d. Since they are on an equal footing in 5d,
one might wonder if the map between them actually corresponds to an
exchange among tensors and vectors in 6d, in which case 5d equivalence
would not lift to one in 6d. That this does not happen can be easily
argued. The fields in the (11,11) example are a proper subset of those
in (19,19), and the equivalence in the former (between A and A$'$)
lifts, as we have just seen, to an equivalence in 6d. This means that
the tensors in the (19.19) example (B and B$'$) do get
identified with each other under our dualities, and there is no
possibility of an interchange with vectors. While this chain of
arguments supports the equivalence in 6 dimensions, a more direct 
demonstration will be given below.

It is perhaps worthwhile to stress that the equivalence we have
demonstrated, between F-theory on orbifolds and M-theory on
orientifolds, is completely different from the well-known relation
betwen F-theory on an orbifold (or any elliptically fibred manifold)
and M-theory on the {\it same} orbifold or manifold, which leads to a
theory in one lower dimension. 

Continuing on the above lines, one can also, in 5 dimensions, map the
models of interest to orientifolds of type IIB. Performing a T-duality
in the 10 direction on the last line of Eq.\chain, we find that the
(51,3) case (models C and C$'$) is also equivalent to type IIB on the
orientifold $T^5/\{1, I_{6789},$ $\Omega,$ $\Omega I_{6789}\}$ which at
first sight is the circle compactification of the orientifold studied
by Gimon and Polchinski\gimpol. We will explain, after a direct
demonstration of the equivalence in 6d, why this is not the case.

We have seen that conjectured F-theory, M-theory and IIB orientifold
duals can be identified by a sequence of T and S-dualities in 5
dimensions.  One may ask if the identifications can be demonstrated
directly in 6 dimensions. Since F-theory has no S-duality, all
dualities having been converted to T-dualities, a chain of
equivalences in 6d is likely to be simpler and more direct. For this,
we need to map the orbifolding symmetries of F-theory directly to
those of type IIB. Just as for M-theory, an inversion in the 11
direction was identified to the operator $\mofl$ in type IIA, for
F-theory a pair of inversions in the 11,12 directions can be
identified with the product $\Omega\mofl$ in type IIB\senorbf. This
follows again by carefully examining the action of these inversions on
the states of type IIB. One support for this identification comes from
the fact that, for orbifold groups which are powers of $Z_2$,
S-duality in type IIB exchanges $\Omega$ and $\mofl$, and in F-theory
this is manifested as the exchange of the 11 and 12 directions.

With this addition to the previous rules, we find directly that
F-theory on the (51,3) (model C) to 6 dimensions can be written as
type IIB on $T^4/\{1, I_{78910},$ $\Omega\mofl I_{78},$ $\Omega\mofl
I_{910}\}$.  We now have a short chain of equivalences:
\eqn\chaintwo{
\eqalign{
&{\rm IIB~on}~T^4/\{1, I_{78910}, \Omega\mofl I_{78}, \Omega\mofl 
I_{910}\}\cr
{\buildrel T_{7}\over\longrightarrow}~ &
{\rm IIA~on}~T^4/\{1, \mofl I_{78910},\Omega\mofl I_{8}, 
\Omega I_{7910}\}\cr
{\buildrel T_{8}\over\longrightarrow}~ &
{\rm IIB~on}~T^4/\{1, I_{78910}, \Omega , \Omega I_{78910} \} \cr} }

The middle line can be recognised precisely to be the M-theory
orientifold on model C$'$ after using the lorentz invariance among the
11 directions to interchange $x^{11}$ and $x^{8}$. This is in
consonance with the fact that we would not expect the M-theory
orientifold to be perturbatively T-dual to the F-theory
compactification. Interchanging $x^{11}$ and $x^{8}$ is a
non-perturbative operation in string theory.  The last line, which is
the result of T-dualizing the F-theory on directions 9 and 10
simultaneously, leads once again (apparently) to the Gimon-Polchinski
model. But clearly the GP model spectrum (with 1 tensor) does not
agree with that of F-theory on the (51,3) (with 17 tensors).

Here we find support for the idea, proposed in the previous section,
that there are actually two distinct versions of model C (in F-theory)
or model C$'$ (in M-theory).  It is now
understood\blpssw\ that the Gimon-Polchinski model is
equivalent to an F-theory compactification on the (3,243) Calabi-Yau,
which in turn is known\mvone\ to be dual to the $E_8\times E_8$
heterotic string on K3 with the symmetric (12,12) embedding of
instanton numbers in the two $E_8$ factors. Thus, on relating our C
and C$'$ models to orientifolds, we have apparently found a second
realisation of these models. 

However, it has recently been noticed\polkthree\ that there is
actually an ambiguity in the action of the orientation-reversal
operator $\Omega$ on twisted-sector states. The model is completely
characterised only after this action is specified.  We
expect that the twofold ambiguity that we discovered above, in models
C and C$'$, will be mirrored in the orientifold construction. Both
models should have the common feature of having both 5- and
9-branes. This is easily seen to arise on T-duality from the two sets
of 7-branes which are wrapped around $x^{7},x^{8}$ and $x^{9},x^{10}$
respectively.

Continuing with the original discussion on the relation to
orientifolds, we note that the same short chain maps the (19,19)
(models B,B$'$) first to M-theory on $(K3\times S^1)/Z_2$ and then to
the orientifold IIB on $T^4/\{1, I_{78910},$ $\Omega I_{78910}S_{10},$
$\Omega S_{10}\}$ studied in Ref.\dpone. (Here, $S_{10}$ is a shift by
a half unit in the 10 direction.) As another consistency in this
mapping we note that we originally had 7-branes wrapped around
$x^{7},x^{8}$ in the F-theory. T-dualising in these directions gives
us the 5-branes of Ref.\dpone\ distributed in exactly the way one
obtains $SO(8)^2$ enhanced gauge group with hypermultiplets in the
adjoint in their picture. We also see that there are no 9-branes, as
expected.

For the (11,11) (models A,A$'$) the situation is slightly
different. There, the analogue of the first line in Eq.\chaintwo\ is
type IIB on $T^4/\{1, I_{78910},$ $\Omega\mofl
I_{78}S_{8}S_{10},$ $\Omega\mofl I_{910}S_8 S_{10}\}$.  M-theory on A$'$
follows as usual after a $T_{7}$.  But this time, we have a shift in
the direction 8 where we want to perform a T-duality. Under $T_{78}$,
the shift $S_{8}$ will be replaced by a ``winding shift'' ${\hat
S}_{8}$ and so we will find the orientifold of type IIB on $T^4/\{1,
I_{78910},$ $\Omega {\hat S}_{8}S_{10},$ $\Omega I_{78910}{\hat
S}_{8}S_{10}\}$. This appears to be a new orientifold of type
IIB. Note that since there were no 7-branes in the F-theory, there are
neither 5-branes nor 9-branes in the T-dual theory. This feature,
reflecting no net tadpole in the one loop worldsheet contributions,
was also present in the $T^4/Z_4$ orbifold of\gjone,\dptwo.

The success of these manipulations encourages us to investigate the
case of F-theory compactifications on 8-dimensional orbifolds down to
4d, and relate them to orientifolds of M-theory on $T^7/(Z_2)^3$,
obtaining N=1 supersymmetric theories in 4 dimensions. We will also
find realizations of these models as type IIB orientifolds on $T^6$.
A larger class of models arises by starting with non-toroidal
orbifolds in F-theory, of the form $(K3\times T^4)/(Z_2)^2$, which we
identify with M-theory on seven-dimensional orientifolds $(K3\times
T^3)/(Z_2)^2$ and IIB on six-dimensional orientifolds $(K3\times
T^2)/(Z_2)^2$.
 
\newsec{Compactifications to four dimensions}

The Calabi-Yau threefolds A, B and C that we have discussed are
special cases of the so-called Voisin-Borcea models, all of which are
defined by taking $(K3\times T^2)/Z_2$, where the $Z_2$ acts as
reflection on the 2-torus and as an involution\ref\nikulin{\NIKULIN}
on K3 which reverses the sign of the $(2,0)$ form. The possible such
involutions have been classified and are labelled by three integers
$(r,a,\delta)$. (For a table of the allowed cases, see for example
Ref.\mvtwo.) One might also imagine constructing a large class of
Calabi-Yau 4-folds of the form $(K3\times K3)/Z_2$ using these
involutions. Indeed, Calabi-Yau 4-folds of this form with $SU(4)$
holonomy have been considered by Borcea\borcea, with the $Z_2$ action
on the K3's labelled by $(r_1,a_1,\delta_1; r_2,a_2,\delta_2)$.  These
involutions reverse the individual $(2,0)$ forms while preserving the
$(4,0)$ form. These therefore provide a large class of $N=1$ vacua in
four dimensions when utilised for F-theory compactifications.

The A,B and C threefolds that we studied in previous sections
corresponded to choosing the three involutions on $K3$ that can be
defined in a toroidal orbifold limit. (By abuse of notation, we
label the involutions A,B and C as well.) In terms of $(r,a,\delta)$
they are (10,10,0), (10,8,0) and (18,4,0) respectively.)  For
fourfolds, there is a variety of orbifold limits that one can
discuss.  We first consider a couple of families which realise a
$(T^4/Z_2 \times K3)/Z_2$ limit and later, some special cases
admitting a $T^8/(Z_2)^3$ orbifold point.

In the former category, let us start with the $Z_2$ acting as a
$(C,\sigma)$ involution on the $T^4/Z_2$ and $K3$ respectively. Here
$\sigma \equiv\sigma(r,a,\delta)$ is a general $K3$ involution
reversing the (2,0) form. Choosing the fibre for F-theory
compactification in the $T^4$ (with coordinates $x^9\ldots x^{12}$;
the $K3$ will be labelled by $x^5 \ldots x^8$) the action of the
$Z_2$'s can be summarised as
\bigskip
\centerline{Model (C,$\sigma$):}\nobreak
\medskip
\begintable
|$x^{12}$|$x^{11}$|$x^{10}$|$x^9$|$x^8\ldots x^5$\elt
$\alpha$|$-$|$-$|$-$|$-$|$+$\elt
$\beta$|$+$|$+$|$-$|$-$|$\sigma$\elt
$\alpha \beta$|$-$|$-$|$+$|$+$|$\sigma$
\endtable

The Hodge numbers and Euler characteristics of this family of 4-folds
can be worked out using the general expression in Ref.\borcea, giving
\eqn\chodge{
\eqalign{
& H^{1,1} = 26 +5r -4a,\qquad H^{2,1} = 88 -4r -4a \cr
& H^{3,1} = 22 -r,\qquad H^{2,2} = 60 +24r -8a \cr
& \chi = 2(2 + H^{1,1} + H^{3,1} - 2 H^{2,1}) + H^{2,2} = 48(r-4)\cr} }
We note that the Euler characteristic $\chi$ is a multiple of 24, as
expected\svw, and is negative for $0 \leq r \leq 3$. Thus,
these cases are likely to lead to F-theory compactifications which are
inconsistent because of tadpoles\svw. For $r=4$ we have no tadpoles,
while for $r>4$ the tadpoles can be cancelled by assuming condensation
of $2(r-4)$ IIB 3-branes in the vacuum.

The equivalence to M-theory follows as before: The above orbifold 
can be written as type IIB on $(T^2\times K3)/\{1, 
\Omega\mofl I_{910},$ $I_{910}\sigma,$ $\Omega\mofl \sigma\}$. 
Our short chain now reads:
\eqn\chainthree{
\eqalign{  
&{\rm IIB~on}~(T^2\times K3)/\{1, \Omega\mofl I_{910}, I_{910}\sigma ,
\Omega\mofl \sigma\}\cr {\buildrel T_{9}\over\longrightarrow}~ & {\rm
IIA~on}~(T^2\times K3)/\{1, \Omega I_{10}, \mofl I_{910}\sigma ,
\Omega\mofl I_{9}\sigma \}\cr
{\buildrel T_{10}\over\longrightarrow}~ &
{\rm IIB~on}~(T^2\times K3)/\{1, \Omega , I_{910}\sigma, 
\Omega I_{910}\sigma \} \cr} }

The middle line, after interchanging $x^{11}$ and $x^{10}$, is
precisely M-theory on the orbifold $S^1/Z_2 \times (T^2\times
K3)/\{1,I_{910}\sigma \}$.  Note that $(T^2\times
K3)/\{1,I_{910}\sigma \}$ are the Voisin-Borcea 3-folds
$CY_{\sigma}$ with Hodge numbers
\eqn\vbhodge{
H^{1,1} = 5 +3r -2a, \qquad H^{2,1} = 65 -3r -2a }
The last line of \chainthree\ is the orientifold of type IIB on
$CY_{\sigma} /\{1, \Omega \}$, which at first sight seems to be the
same as Type I on $CY_{\sigma}$. The subtleties that arose in the 6d
examples recur, however, in this case as well, and we will have more
to say on this below.

What can one say about the spectrum of these models? Unfortunately,
F-theory on 4-folds has not so far been studied in great detail. So we
restrict ourselves to some naive deductions. We can learn something
about the gauge group by looking at the singularity of the fibre. The
fibre degenerates to $T^2/Z_2$ over certain fixed divisors in the
base. These consist of the following: The element $\alpha$ has 4
copies of $K3/\sigma$ as its fixed locus. The element $\gamma$ also
has degenerating fibres over the fixed loci of $\sigma$ in K3. These
latter are known to consist of $(k+1)$ components (see for instance
Ref.\mvtwo) where $k=\half(r-a)$. Of these, $k$ are rational curves and
1 is of genus $g=\half(22 -r -a)$ Since each component contributes an
$SO(8)$ (at least naively), one has apparently an $SO(8)^{k+5}$ gauge
group. An exception is the $(C,A)$ case, since the Enriques involution
$A$ has no fixed points and therefore the group is $SO(8)^4$.  This
also seems consistent with the following fact: assuming no $U(1)$
factors, we must have
\eqn\rank{
r(V) = H^{1,1}(X) -H^{1,1}(B) -1}
Examining the table for $(C,\sigma)$, we see that
$H^{1,1}(B)=H^{1,1}(CY_{\sigma})$. Using \vbhodge\ and \chodge\ we
obtain $r(V)=4(k+5)$ as desired. In the IIB orientifold, these vector
multiplets come from the open string twisted sector states.

The closed string sectors (twisted and untwisted), 
of the orientifold give rise to some
neutral chiral multiplets. These are easy to count. They are simply
the closed string states of Type I on $CY_{\sigma}$ and are
$H^{1,1}+H^{2,1}+1$ in number. (The 1 counts the dilaton chiral
multiplet.) In the case of our 3-folds, Eqn.\vbhodge\ determines this
number to be $71-4a$. 

Finally, there can be charged chiral multiplets as well. These are
determined by the nature of the fixed divisor on the base. We will
assume that the only charged matter comes from the divisor containing
the genus $g$ curve. This was also the source of charged matter in 6
dimensions. Then there are $g$ chiral multiplets in the adjoint of the
$SO(8)$ associated to this divisor. Thus we have $24g$ multiplets
charged under the gauge group, and $4g$ which are neutral. We will show
that this spectrum satisfies a non-trivial consistency requirement.

Consider M-theory compactified on the same 4-fold down to 3 spacetime
dimensions. The number of neutral multiplets in 3d (where vectors are
dual to scalars) is $H^{1,1} +H^{2,1} +H^{3,1}$. Thus one has the
following constraint from compactification of our original theory on a
circle where the gauge group is Higgsed:
\eqn\mtheory{
r(V) +c^0 +1 = H^{1,1} + H^{2,1} + H^{3,1} }
where $c^0$ is the number of neutral chiral multiplets, and the 1 comes
from the gravity multiplet. Substituting $r(V) =4(k+5) =2r -2a -20$
and $c^{0} =4g +(71 -4a) =115 -2r -6a$, together with the hodge
numbers from \chodge, we see that Eqn.\mtheory\ is satisfied. This
gives us confidence that our assumptions were well-founded. There can
be extra contributions from the $\chi/24 =2r -8$ 3-branes of IIB
necessary to cancel the tadpole. These will be invisible in
\mtheory\ since the M-theory compactification would also have extra
membranes.

One can similarly define another family of fourfolds admitting
$(T^4/Z_2 \times K3)/Z_2$ orbifold limits. Here the $Z_2$
acts as a $(B,\sigma)$ involution and one can consider F-theory
compactification on 
\bigskip
\centerline{Model (B,$\sigma$):}\nobreak
\medskip
\begintable
|$x^{12}$|$x^{11}$|$x^{10}$|$x^9$|$x^8\ldots x^5$\elt
$\alpha$|$-$|$-$|$-$|$-$|$+$\elt
$\beta$|$+$|$+$|$-,\half$|$-$|$\sigma$\elt
$\alpha \beta$|$-$|$-$|$+,\half$|$+$|$\sigma$
\endtable
This time the Hodge numbers are 
\eqn\bhodge{
\eqalign{
& H^{1,1} = 12 +2r -a,\qquad H^{2,1} = 24 -2a \cr
& H^{3,1} = 52 -2r -a,\qquad H^{2,2} = 252 -4a \cr
& \chi = 288 \cr } }

The same manipulations as in \chainthree\ now give us M-theory on the
orbifold $(S^1/Z_2 \times T^2 \times K3)/Z_2$ where the second $Z_2$
acts as a pure shift $+ \half $ on the $S^1$ and the usual
$\{I_{910}\sigma\}$ on the $T^2 \times K3$. However, after another
T-duality we end up with a IIB orientifold with a `winding shift'.

The gauge group in this case is $SO(8)^2$ at the orbifold point,
independent of $\sigma$, for exactly the same reasons as in the 6d
model B: (19,19). The charged matter in this case consists of 2
adjoints of $SO(8)$ which Higgs the group at a generic point giving
$U(1)^8$ together with 8 chiral multiplets.  More neutral chiral
multiplets come, as before from the closed string sector states of the
orientifold, which once again turn out to be $71-4a$ in number. It is
again a non-trivial check to see that \mtheory is
satisfied. Substituting $r(V)=8$,$c^0 =8 +(71-4a)$ and the hodge
numbers from \bhodge, we once again match the expressions.

The number of 3-branes needed to cancel the 2 form tadpole is always
$\chi /24 =12$.  We also note in passing that F-theory on the $(C,B)$
orbifold is not the same as on the $(B,C)$ --- they have different
gauge groups, $SO(8)^6$ and $SO(8)^2$ respectively. Thus it does
matter whether we choose our fibre to have either the $C$ or $B$
involution, and reflects a lack of 12 dimensional Lorentz
invariance.

F-theory compactifications on many other Borcea 4-folds can also be
considered, and we reserve this for future study. For instance,
another interesting family is the $(A,\sigma)$ all of whose members
have $\chi =288$. We proceed instead to examine a few special cases,
some of which are members of the two families described above. These
are the cases where one can go to the $T^8/(Z_2)^3$ point in moduli
space.

\newsec{Some Particular Four-dimensional Compactifications}

The $T^8/(Z_2)^3$ orbifold point of $(K3\times K3)/Z_2$ is realised in
6 distinct fourfolds labelled by their involutions -- (A,A), (B,B),
(C,C), (A,B), (A,C) and (B,C). We note once again that each of these
manifolds can realise a number of different F-theory vacua depending
on the choice of fibre.

Since it is tedious to list all 7 nontrivial elements of $(Z_2)^3$, in
this section we will often list only the three generators $\alpha$,
$\beta$ and $\gamma$, one for each $Z_2$ factor. It must be kept in
mind that, depending on the shifts included, the other elements that
we do not always write explicitly may or may not have fixed points.

Let us focus on the (C,C) fourfold, which has no shifts. This is, of
course, a member of the family studied in the previous section with
$\sigma =C =(18,4,0)$ and is the 4d analogue of the threefold (51,3).
Its generators are as follows:
\bigskip
\centerline{Model (C,C):}\nobreak
\medskip
\begintable
|$x^{12}$|$x^{11}$|$x^{10}$|$x^9$|$x^8$|$x^7$|$x^6$|$x^5$\elt
$\alpha$|$+$|$+$|$+$|$+$|$-$|$-$|$-$|$-$\elt
$\beta$|$-$|$-$|$-$|$-$|$+$|$+$|$+$|$+$\elt
$\gamma$|$-$|$-$|$+$|$+$|$-$|$-$|$+$|$+$
\endtable

This corresponds to type IIB on the orientifold $T^6/\{1, I_{5678},$
$\Omega\mofl I_{910},$ $\Omega\mofl I_{78}\}$. (Recall that here we are
listing only the generators and not all the nontrivial elements.)  
By a T-duality in the 9 direction we get IIA on $T^6/\{1, I_{5678},$
$\Omega I_{10},$ $\Omega\mofl I_{789}\}$ which is the same, after 
interchange of $x^{10}$ and $x^{11}$, as the M-theory orientifold
$T^7/(Z_2)^3 = S^1/Z_2 \times T^6/(Z_2)^2$ which we label (C$'$,C$'$):
\bigskip
\centerline{Model (C$'$,C$'$):}\nobreak
\medskip
\begintable
|$x^{11}$|$x^{10}$|$x^9$|$x^8$|$x^7$|$x^6$|$x^5$\elt
$\alpha$|$+$|$+$|$+$|$-$|$-$|$-$|$-$\elt
$\beta$|$-$|$+$|$+$|$+$|$+$|$+$|$+$\elt
$\gamma$|$+$|$-$|$-$|$-$|$-$|$+$|$+$
\endtable
Applying, instead, a T-duality in the pair of directions 9,10 we find,
after a change of generators, an orientifold of type IIB on
$T^6/\{1, \Omega,$ $I_{5678},$ $I_{78910}\}$ which is, in appearance, the
same as the model studied in Ref.\berkl.

Let us look at the spectrum of this model. The Hodge numbers and
Euler characteristic of the fourfold are given by
\eqn\hodgeeuler{
\eqalign{
(H^{1,1},H^{2,1},H^{3,1},H^{2,2}) &= (100,0,4,460)\cr
\chi &= 672\cr} }
{}From the results of the previous section we know the following. The
gauge group is $SO(8)^{12}$, corresponding to $k=7$, which can also be
argued as in the case of the (51,3). Since $g=0$ in this case, there
are apparently no charged hypermultiplets to Higgs the gauge
group. The number of neutral chiral multiplets, $71 -4a$ is therefore
55, which is also the number of chiral multiplets obtained in the
closed string sector (twisted and untwisted) in \berkl. However, the
open string spectrum is different from theirs. This is once again due
to the ambiguity that we encountered in the (51,3) case. We expect that
an orientifold with the complementary action of $\Omega$ on twisted
sector states will realise the spectrum that we find here. There is no
contradiction between this and the fact that the closed string twisted
sector states of the two theories agree. In fact, the closed string
twisted sector states are in $N=2$ hypermultiplets (each of which
decomposes to two $N=1$ chiral multiplets) before the $\Omega$
projection. We therefore see that both actions of $\Omega$, in this 
sector, will produce chiral multiplets.

This can be understood geometrically in F-theory, where we can
immediately identify two distinct Calabi-Yau manifolds to which this
orbifold can be smoothened. Changing the K\"ahler structure (by
blow-ups) leads to the Borcea 4-fold which we have been discussing.
The other way to go away from the orbifold limit is to think of the
singular base $T^6/(Z_2)^3$, which is actually $(T^2/Z_2)^3$, as a
limit of $(P^1)^3$ as one varies the complex structure.  Thus, it is
the limit of an elliptic fibration of the form
\eqn\ellipfib{
y^2 = x^3 + f(u,v,w)x + g(u,v,w) }
where $u,v,w$ are the inhomogeneous coordinates on $(P^1)^3$ and
$(f,g)$ are polynomials of degrees (8,12) respectively, in all their
arguments.  It is easy to see that this manifold will have $H^{1,1}=4$
and $H^{3,1}=2916$, the latter number arises as $(9)^3 + (13)^3 -
3-3-3-1$ from the polynomial deformations in Eq.\ellipfib\ above.

This smooth manifold, fibred over a $(P^1)^3$ base, is the strict
analogue of the (3,243) Calabi-Yau 3-fold that has been studied in the
context of F-theory compactification to 6 dimensions. Thus, we claim
that the $Z_2\times Z_2$ orientifold of type IIB on $T^6$ studied in
Ref.\berkl\ is dual to F-theory on the above elliptically fibred
Calabi-Yau 4-fold. The relation between the two follows from a rerun
of the arguments above at the orbifold point.  We note that this
manifold posseses many interesting features, such as the symmetry
under the interchange of the $P^1$'s. The corresponding exchange
symmetry had reflected itself in the heterotic dual of the (3,243)
case in 6 dimensions as strong-weak duality. We expect a similar
effect here, leading to a heterotic-heterotic dual family in 4d
related by the permutation group $S_3$.

It will be interesting to see if there are still other distinct ways
to smoothen this orbifold, since after all 4-folds have other
deformations besides those of K\"ahler and complex structure. Indeed,
there are actually 4 ways to project out an N=1 multiplet from an N=4
multiplet since the latter, which is necessarily a vector multiplet,
decomposes into one vector and three chiral multiplets of N=1.  This
is in tantalising correspondence with the fact that a Calabi-Yau
4-fold has 4 independent Hodge numbers,
$(H^{1,1},H^{2,1},H^{3,1},H^{2,2})$.  Also notice that the discrete
torsion for an orbifold group of the form $(Z_2)^3$ takes values in
$Z_2\times Z_2$, which is of order four.

An argument that we will outline in the case of M-theory indicates
that we will in general encounter the full discrete torsion. The
fourfold ambiguity is also present in the IIB orientifold on
$T^6/(Z_2)^2$ since in addition to the ambiguity in $\Omega$ one could
also introduce the conventional discrete torsion in the conformal
field theory.

Now let us say a few words about the (C$'$, C$'$) compactification of
M-theory. Of the 7 nontrivial elements of the $(Z_2)^3$ orbifolding
group, 1 element ($\beta$) corresponds to inversion of a single space
coordinate, 3 elements ($\alpha$, $\gamma$ and $\alpha\gamma$)
correspond to inversions of 4 space coordinates, and the remaining 3
($\alpha\beta$, $\beta\gamma$ and $\alpha\beta\gamma$) correspond to
inversions of 5 space coordinates. As we know, each element of the
last set produces 16 M-theory 5-branes. Thus we have a total of 48
M-theory 5-branes in the vacuum, each of which appears as a 3-brane in
spacetime, as two of its dimensions are wrapped around an internal
$T^2$. 

The actual degrees of freedom carried by these branes will depend on
what is projected out from each twisted sector upon requiring
invariance under the remaining elements. Before projection, the brane
multiplets are gauge multiplets of $N=4$ supersymmetry in 4d, thus
their content is $(A_\mu, 6\phi)$. Projection will select from this an
$N=1$ multiplet, which can be either a vector or a chiral
multiplet. 

Similarly, the other 4 twisted sectors will each produce a rank-16
gauge bundle, for which the actual gauge group and the multiplets
projected out/in have to be determined. Our present knowledge 
is not yet powerful enough to work this out directly in M-theory
and in four
dimensions we do not have the constraint of satisfying gravitational
anomaly cancellation equations. Going by analogy with the C$'$ model of
M-theory on $T^5/(Z_2\times Z_2)$ and what we have seen
in the equivalent description in terms of 
IIB orientifolds and F-theory, it seems quite certain that there
will be at least two distinct consistent projections for this model
in the framework of M-theory. 

In fact, we can put forward the following argument in M-theory that
these different projections are indeed the manifestation of discrete
torsion. In a $(Z_2)^3$ orbifold of M-theory, we could imagine turning
on discrete torsion in a $(Z_2)^2$ which would be seen in string
theory (for instance, the $(Z_2)^2$ in the second factor of $S^1/Z_2
\times T^6/(Z_2)^2$). But Lorentz invariance in 11 dimensions can
turn the action of these spacetime $Z_2$'s into those with an
$\Omega$. This therefore turns what was discrete torsion into a
complementary projection in some $\Omega$ twisted sector. A T-duality
ensures that this ambiguity exists in the orientifolds of IIB as
well. We will see a realisation of this in an example in the next section.
It is also clear that in general, all the discrete torsion
elements will come into play, indicating a fourfold ambiguity in the
definition of our orbifolds/orientifolds.  This would come from a
twofold ambiguity in $\Omega$ and a twofold conventional discrete
torsion.

Finally, for the sake of completeness we list the remaining
$T^8/(Z_2)^3$ orbifolds of F-theory. Since the generators $\alpha$ and
$\beta$ are the same as in the $(C,C)$ case, we only need to list
$\gamma$ individually for each of the cases.
\bigskip
\centerline{The element $\gamma$ for toroidal fourfolds}\nobreak
\medskip
\begintable
|$x^{12}$|$x^{11}$|$x^{10}$|$x^9$|$x^8$|$x^7$|$x^6$|$x^5$\elt
(A, A)|$-,\half$|$-$|$+,\half$|$+$|$-,\half$|$-$|$+,\half$|$+$\elt
(B, B)|$-$|$-$|$+,\half$|$+$|$-$|$-$|$+,\half$|$+$\elt
(B, A)|$-$|$-$|$+,\half$|$+$|$-,\half$|$-$|$+,\half$|$+$\elt
(B, C)|$-$|$-$|$+,\half$|$+$|$-$|$-$|$+$|$+$\elt
(C, A)|$-$|$-$|$+$|$+$|$-,\half$|$-$|$+,\half$|$+$\elt
(C, B)|$-$|$-$|$+$|$+$|$-$|$-$|$+,\half$|$+$
\endtable
The corresponding M-theory orientifolds are realised as $T^7/(Z_2)^3$
where again the elements $\alpha$ and $\beta$ are the same as in the
(C$'$, C$'$) case defined above, so we only need to list the element
$\gamma$. 
\bigskip
\centerline{The element $\gamma$ for 7d orientifolds}\nobreak
\medskip
\begintable
|$x^{11}$|$x^{10}$|$x^9$|$x^8$|$x^7$|$x^6$|$x^5$\elt
(B$'$, B$'$)|$-$|$+,\half$|$-$|$-$|$-$|$+,\half$|$+$\elt
(B$'$, C$'$)|$+.\half$|$-$|$-$|$-$|$-$|$+$|$+$\elt
(B$'$, A$'$)|$-$|$+,\half$|$-$|$-,\half$|$-$|$+,\half$|$+$\elt
(C$'$, A$'$)|$-$|$+$|$-$|$-,\half$|$-$|$+,\half$|$+$\elt
(C$'$, B$'$)|$-$|$+$|$-$|$-$|$-$|$+,\half$|$+$
\endtable

It should be clear that T-duality transformations in the 9 direction
produce mappings between each pair of F and M-theory compactifications
listed above. A T-duality in the 9 and 10 directions takes us to type
IIB orientifolds, none of which have been independently invesigated.

\newsec{Models With Higher Supersymmetry in Four Dimensions}

Although this is somewhat off the main point of this paper, we find it
appropriate to discuss F-theory and M-theory compactifications to 4
dimensions with $N>1$ supersymmetry. The first example is a rather
simple one: F-theory on $T^8/Z_2$, where the nontrivial element of
$Z_2$ acts by reversing all the 8 directions $x^5,\ldots,x^{12}$. The
orbifold $T^8/Z_2$ cannot be smoothened to a Calabi-Yau, nevertheless
it is a consistent manifold for string compactification. In
Ref.\dmone, it was proposed that IIB on this orbifold is dual to
M-theory on $T^9/Z_2$, while in Ref.\ref\senorbdual{\SENORBDUAL}\ it
was argued that M-theory on this orbifold has 16 M-theory 2-branes in
3 dimensions. This shows that the appropriate analogue of $\chi/24$,
the number of branes required to cancel tadpoles, is 16 in this case.

The same argument thus implies that F-theory on $T^8/Z_2$ has 16
3-branes in 4 dimensions. The resulting theory is N=4 super-Yang-Mills
theory, with gauge group $U(1)^{16}$ coming from separated 3-branes,
along with further contributions from the untwisted sector. Each
3-brane carries the complete N=4 gauge multiplet with 6 collective
coordinates for its motion along $x^5,\ldots,x^{10}$.

By the dualities discussed above, we can relate this to M-theory on
$T^5/Z_2\times T^2$. In this picture, the theory has 16 M-theory
5-branes wrapped on $T^2$ to form 3-branes. A further compactification
on a circle to 3 dimensions leads to a theory with 16 5-branes wrapped
on $T^3$ to form 2-branes. This is electric-magnetic dual to the
compactification of M-theory directly on $T^8/Z_2$ discussed in
\senorbdual, where one has 16 fundamental M-theory 2-branes rather
than wrapped 5-branes. Thus the resulting picture is perfectly
consistent.

A more interesting example is F-theory on $T^8/(Z_2\times Z_2)$ where
the orbifolding group is defined to be generated by the elements
$\alpha$ and $\beta$ that we used in defining the (C,C) model in the
previous section and thus actually is $T^4/Z_2 \times T^4/Z_2$. 
This becomes the orientifold of type IIB on
$T^6/\{1, I_{5678},$ $\Omega\mofl I_{910},$ $\Omega\mofl
I_{5678910}\}$. A T-duality in the 9 direction maps this to type IIA
on $T^6/\{1, I_{5678},$ $\Omega I_{10},$ $\Omega I_{567810}\}$.
Interchanging the 10 and 11 directions gives M-theory on
$T^7/(Z_2\times Z_2)$ where now the group is defined through $\alpha$
and $\beta$ of the (C$'$, C$'$) model. Clearly this is just M-theory
compactified on model C$'$ to 6 dimensions, and then further on $T^2$
to 4 dimensions. By another pair of T-dualities, this is also
equivalent to type IIA on the orbifold limit of the $(51,3)$
Calabi-Yau 3-fold, to 4 dimensions, with N=2 supersymmetry. This last is 
precisely the case where discrete torsion in string theory was 
studied\fiqdisc\vwdisc\. We will see that this ambiguity can be 
mapped onto an orientifolding ambiguity. For that we note that a
T-duality ($T_{56}$) takes the original IIB orientifold into IIB on
$T^6/\{1, I_{5678}, \Omega I_{56910}, \Omega I_{78910} \}$.

Instead let us start out with the particular $T^7/(Z_2)^2$ orientifold of 
M-theory, which, in IIA language, would be IIA on 
$T^6/\{1, \mofl I_{78910}, I_{5678}, \mofl I_{56910} \}$. 
We immediately see that $T_{10}$ gives us IIB on the (51,3) Calabi-Yau.
In the absence of discrete torsion, this is distinct from IIA on this
orbifold, which we just encountered. However, turning on discrete
torsion in the IIA theory makes it the same as the IIB theory (or vice
versa).  We will now map this M-theory/IIA orientifold into what is
apparently the same IIB orientifold in the last line of the previous
paragraph.  To do so, we use the Lorentz invariance of M-theory to
interchange directions 5 and 11. This obtains for us the orientifold,
IIA on $T^6/\{1, \Omega I_{578910}, \Omega\mofl I_{678}, \mofl
I_{56910} \}$.  Finally, $T_5$ results in IIB on $T^6/\{1, \Omega
I_{78910}, \Omega I_{5678}, I_{56910} \}$, which is the claimed
orientifold (after a relabelling of 7,8 as 9,10). Thus we see that the
ambiguity of discrete torsion in string theory on the (51,3) orbifold
has manifested itself in an ambiguity in the (equivalent) orientifold
of IIB on $T^6/\{1, I_{5678}, \Omega I_{56910}, \Omega I_{78910}
\}$. The two different projections give us either IIA or IIB on the
(51,3).

One can similarly generate N=2 compactifications of M and F-theory to
4 dimensions which correspond to type IIA on the (11,11) or (19,19)
Calabi-Yau 3-folds, by picking a suitable $Z_2\times Z_2$ subgroup of
the $(Z_2)^3$ group of the previous section for the (C,A) and (C,B)
models.

Although this discussion concerns compactifications with higher
supersymmetry, it is worth noting here for completeness that there is
an orbifold with lower supersymmetry than N=1, obtained by starting
with model (C,C) of the previous section and adjoining the generator
$\delta$ with acts as $(+,-,+,-,+,-,+,-)$\foot{The contents of this
paragraph were essentially explained to us by E. Witten.}. This is
precisely the $T^8/(Z_2)^4$ orbifold which leads to the Joyce
8-manifold of spin(7) holonomy\ref\joyce{\JOYCE}. The extra generator
kills another half of the supersymmetry, leading apparently to an
N=$\half$ compactification of F-theory.  Unfortunately, as is evident
from the structure of $\delta$, the action on the fibre (the first two
coordinates) is not in SL(2,Z), so this is not a valid
compactification of F-theory at all, avoiding a potential paradox. It
is, however, a valid orbifold for M-theory to three dimensions, where
it leads to N=1 supersymmetry\acharyaone.

\newsec{Discussion and Conclusions}

We have defined a class of orbifolds of F-theory and, in one-to-one
correspondence, a class of orientifolds of M-theory, which lead to N=1
supersymmetry in 6 and 4 spacetime dimensions. The totality of
available examples in the latter class is of the order of a few
hundred. From the F-theory point of view, the examples are all of the
form $(K3\times T^2)/Z_2$ (to six dimensions) and $(K3\times K3)/Z_2$
(to four dimensions). 

We have been able to identify the spectra of these models in a number
of cases, and shown that they pass various consistency checks. One
relatively surprising discovery is the presence of an ambiguity,
analogous to discrete torsion, in those models where the orbifolding
group has a sufficient number of elements with fixed points. In the
six-dimensional case, we related this to the possibility of defining
two distinct actions of the orientation-reversal operator $\Omega$ in
orientifold models. In four dimensions we predict a larger ambiguity,
coming from a combination of conventional discrete torsion and the
ambiguity discussed above. 

This may eventually lead to the discovery of some new property
inherent in M and F theories, for the following reason. F-theory, in
particular, can be interpreted (though not yet with complete success)
as a 12-dimensional theory compactified on a Calabi-Yau $p$-fold. In
the simplest case where the ambiguity appears, it relates F-theory on
two Calabi-Yau 3-folds, with Hodge numbers (51,3) and (3,243), which
do not appear to be related by a corresponding ambiguity in string
theory. The known ambiguity in string theory relevant to the present
case relates instead the (51,3) to the (3,51), and has been
interpreted at least in this special case as mirror symmetry. In
contrast, F-theory cannot be compactified on the (3,51) since the
result does not satisfy 6d anomaly cancellation. We might try to
generalise from the behaviour of 4 dimensional string theories with
discrete torsion at conifold singularities. The resulting theory is
non-singular. The theory without discrete torsion would however have
been singular and we now understand that there are massless particles
and the theory undergoes a phase transition. The analogous phase
transitions in six dimensions are mediated by tensionless
strings. Indeed, the (51,3) can undergo an extremal transition to the
(3,243) from the shrinking of 16 2-cycles in the base. These give rise
to tensionless strings.  Perhaps the presence of the analogue of
discrete torsion prevents the theory from making this transition. This
clearly needs to be studied further.

{}From the point of view of both M and F-theory, it is evident that
once the orbifold group is larger than $Z_2$, one has to find a
precise way to determine how twisted-sector states are projected in or
out by the other elements. Besides the bearing of this question on the
ambiguity discussed above, it is also of interest in that it may help
uncover the basic structure underlying these theories, beyond the
important discovery\witfive\vafaf\ref\seiwit{\SEIWIT} that
twisted-sector states in these theories frequently arise from
condensation of branes.  This discrete-torsion-like behaviour in
M-theory, is another indication, like the appearance of twisted
sectors, that M-theory orbifolds show some similar behaviour to that
of conformal field theories.

A very important point that emerged recently from the work of
Sen\senorbf\ is that although F-theory in an orbifold limit is dual to
a type IIB orientifold, the naive moduli space that would be suggested
by perturbative study of the orientifold is quite different from the
exact moduli space, which is successfully predicted by F-theory in the
case studied. It seems likely that the nontrivial relation between
the two persists in the cases we have considered here, namely
compactifications to 6 and 4 dimensions. (This ultimately suggests that 
F-theory is the most powerful of the various formulations, though 
ironically it is also the one whose foundations are the least 
well-understood. This is also supported by the fact that all 
M-theory orbifolds (at least of the $Z_2$ variety) seem to have a 
realisation in F-theory.) 

Given the relative solvability of orbifold compactifications in string
theory, one may treat the Borcea 4-fold examples that we have
presented as laboratories to study a wide variety of physical effects:
phase transitions\ref\witphasemf{\WITPHASEMF}\ and nonperturbative
superpotentials\ref\witnonpert{\WITNONPERT}\ in particular. We hope to
report on such investigations in the future. The study of F-theory
compactifications to 4 dimensions (which has been initiated in the
above references and in Refs.\svw,\ref\brunschim{\BRUNSCHIM}) is still
in its infancy, and it is likely that it will lead to a new 
understanding of physically interesting string compactifications.

{\bf Note Added:} After the work presented in Sections 2 and 3 was
completed, some preprints appeared which overlap with a few of the
observations contained therein. In particular, Ref.\ref\gjtwo{\GJTWO}\
talks about realizations of some N=1 6d models in terms of M-theory,
F-theory and string theory. Two more papers, Refs.\ref\blumzaf{\BLUMZAF}%
\ref\dpthree{\DPTHREE}, confirm that there exists a consistent
variant of the Gimon-Polchinski model for which $\Omega^2=1$, whose
spectrum corresponds with that of F-theory on the (51,3)
Calabi-Yau. (In the GP model, one has $\Omega^2=-1$ on certain
twisted-sector states.)

\medskip
\noindent{\bf Acknowledgements} 
We are grateful to J. Blum, A. Dabholkar, O.Ganor, D.Gross, P. Horava,
D. Morrison, J. Polchinski, J. Schwarz, C. Vafa, A. Zaffaroni and
especially A. Sen and E. Witten for helpful discussions.  One of us
(S.M.)  acknowledges the hospitality of the Institute for Advanced
Study, Princeton, and the Physics Department at Caltech, where part of
this work was done.

\listrefs   
\bye